\documentclass[prl,aps,twocolumn,showpacs,amsmath,amssymb,floatfix]{revtex4}

\usepackage{graphicx}% Include figure files
\usepackage{dcolumn}% Align table columns on decimal point
\usepackage{bm}% bold math
\usepackage[utf8]{inputenc}
\usepackage[T1]{fontenc}
\usepackage{mathptmx}
\usepackage{}
\begin{document}

\title[Effect of Sr doping on structure, morphology and transport properties of Bi$_2$Se$_3$ epitaxial thin films]{Effect of Sr doping on structure, morphology and transport properties \\of Bi$_2$Se$_3$ epitaxial thin films}
% Force line breaks with \\

\author{S.O. Volosheniuk}
 \altaffiliation[Also at ]{P.N. Lebedev Physical Institute of the RAS, 119991, Moscow, Russia}%Lines break automatically or can be forced with \\
\affiliation{
Skolkovo Institute of Science and Technology, Skolkovo, Moscow Region, 143025,Russia%\\This line break forced with \textbackslash\textbackslash
}%
 \altaffiliation[Also at ]{P.N. Lebedev Physical Institute of the RAS, 119991, Moscow, Russia}%Lines break automatically or can be forced with \\

\author{Yu.G. Selivanov}
 \email{selivan@lebedev.ru}
\author{M.A.Bryzgalov}
\author{V.P. Martovitskii}
\author{A.Yu. Kuntsevich}
 \email{alexkun@lebedev.ru}
 \altaffiliation[Also at ]{ National Research University Higher School of Economics, 101000 Moscow, Russia}%Lines break automatically or can be forced with \\
\affiliation{
P.N. Lebedev Physical Institute of the RAS, 119991, Moscow, Russia%\\This line break forced with \textbackslash\textbackslash
}%

\date{\today}% It is always \today, today,
             %  but any date may be explicitly specified

\begin{abstract}
We report molecular beam epitaxy growth of Sr-doped Bi$_2$Se$_3$ films on (111) BaF$_2$ substrate, aimed to realize unusual superconducting properties inherent to Sr$_x$Bi$_2$Se$_3$ single crystals. Despite wide range of the compositions, we do not achieve superconductivity. To explore the reason for that we study structural, morphological and electronic properties of the films and compare them to the corresponding properties of the single crystals.
The dependence of the c-lattice constant in the films on Sr content appears to be more than an order of magnitude stronger than in the crystals. Correspondingly, all other properties also differ substantially, indicating that Sr atoms get different positions in lattices. We argue that these structural discrepancies come from essential differences in growth conditions.  Our research calls for more detailed structural studies  and novel growth approaches for design of superconducting Sr$_x$Bi$_2$Se$_3$ thin films.

\end{abstract}

\maketitle

\section{\label{Introduction}Introduction}
In past decade three-dimensional topological insulators(3D TI) gain a lot of interest. This kind of quantum materials has insulating bulk and topologically protected conducting surfaces with spin-momentum locking, and, hence nontrivial electrodynamical properties.
Classical layered thermoelectric materials Bi$_2$Se$_3$ and Bi$_2$Te$_3$ appeared to be the most explored among variety of 3D TIs due to well developed technology of synthesis and wide band gap.

Introduction of Cu, Sr, Nb was found to make bismuth selenide superconducting\cite{cusupercondyear,Ando,Ando2,Shen-LI,smylie,Asaba,joaf,KuntsevichNematic,deVisser,pso}. This superconductivity  is attractive both from fundamental point of view (as  its mechanism and possible topological nature are not explored yet) and for applications (i.e. the platform for future quantum computations if the superconductivity is topological).
However, all experimental observations of superconductivity in these materials so far were restricted to bulk single crystals.

Evidently, for any practical applications, scalable thin-films based technology is required, rather than exfoliation of thin single crystalline flakes. Molecular beam epitaxy of parent material Bi$_2$Se$_3$ on various substrates was developed in past few years \cite{Crystals}, record mobilities of $\sim$ 16000 cm$^2/Vs$ were achieved \cite{Ohoho16000}, and possibility of Fermi level tuning both by gate and by doping were demonstrated~\cite{ptype}.
Apparently, the growth of superconducting M$_x$Bi$_2$Se$_3$ (where M = Cu, Sr, Nb) thin films would be  a future great achievement in technology of these topological materials.

As superconductivity was first discovered in the Cu$_x$Bi$_2$Se$_3$ crystals in 2010 \cite{cusupercondyear}, thin films with the same dopant were also grown soon \cite{PRB24075, cufilms3,cufilms2}, and turned out to be non superconductive. Surprisingly, instead of increase  of carrier density,  the epitaxial Cu-doped films  even demonstrated a tendency to insulating behavior \cite{cufilms}.
Bulk Sr$_x$Bi$_2$Se$_3$, discovered in 2015 \cite{joaf}, seems to be more prospective than Cu-doped Bi$_2$Se$_3$, as Sr-doped material is air-stable and demonstrates almost 100\% superconductive volume~\cite{deVisser,KuntsevichNematic,smylie}. In addition, the  bulk carrier density  in Sr$_x$Bi$_2$Se$_3$ ($\sim 2\cdot 10^{19}$ cm$^{-3}$) is an order of a magnitude smaller than in Cu$_x$Bi$_2$Se$_3$. Such low density for  ultrathin ($\sim 10$ nm) film would provide 2D  carrier density per unit area as small as $2\cdot10^{13}$ cm$^{-2}$. It opens a way for switching superconductivity using routinely available field effect transistor structure.

Bi$_2$Se$_3$ films are frequently grown by two step procedure on popular substrates such as (111) Si \cite{vicinalSi111, OhSi111} and (0001) Al$_2$O$_3$ \cite{OhAl2O3, AndoAl2O3}, despite large lattice mismatch (7 to 15\%) relying on Van-der Waals epitaxy. Several studies successfully used MBE growth of Bi$_2$Se$_3$ films on cleaved (111) BaF$_2$ substrates with 5\% in plane lattice misfit employing two-step\cite{oveshnikov} or single step \cite{brasBaF2, SpringholzBaF2} method. Optically transparent and insulating BaF$_2$ substrate is handy for transport and spectroscopic studies. In contrast to Si and Al$_2$O$_3$, it has thermal expansion coefficient (18.7$\times$ 10$^{-6}$ K$^{-1}$) reasonably close to that of Bi$_2$Se$_3$ ($\sim$12 $\times$ 10$^{-6}$ K$^{-1}$)\cite{thermexp}, thus reducing thermal stress. With in-situ deposited BaF$_2$ cap layer\cite{Melnikov}, an optical-friendly (111)BaF$_2$/Sr$_x$Bi$_2$Se$_3$/BaF$_2$ structure would allow to study the optical response of the superconducting condensate in Sr-doped Bi$_2$Se$_3$ film and to conclude on the symmetry of  the superconducting energy gap.

To the best of our knowledge, epitaxial Sr-doped Bi$_2$Se$_3$ films were reported in papers two times\cite{chinawork,ptype}. Ref.\cite{chinawork} presents only two films with $x=0.05$, 0.13 and concentrates on micro structural properties (searching for  Sr atoms positions). In Ref.\cite{ptype}, devoted to (Ca$_x$Bi$_{1-x}$)$_2$Se$_3$ system, several  Sr-doped films with $x\leq 1\%$ are mentioned in supplementary. Thus, the systematic study of growth process and its correlation with structural and transport properties is lacking. Our research aims to fulfill this gap.

We report epitaxial growth, detailed structural, and low-temperature magnetotransport studies of Sr-doped Bi$_2$Se$_3$ thin films on BaF$_2$ substrate. Despite the wide range of growth  parameters, the superconductivity was not achieved. In order to find out the possible reason for that we  compared  structural properties of superconducting bulk crystals and thin films.We observe essential enhancement of the c-axis parameters with $x$ in the films compared to that in bulk crystals. Morphology and transport studies suggest that Sr introduces disorder to the films. Our data thus indicate that Sr atoms in films and bulk crystals take different positions in the lattices. The position of Sr atoms is therefore decisive for superconductivity.

\section{\label{growthandtech}Growth and structural characterization techniques}

\subsection{\label{growth}Growth technique}
Growth of Sr-doped Bi$_2$Se$_3$ films on cleaved ($1\,1\,1$) BaF$_2$ substrates was performed in a MBE system EP-1201 with a pressure $3\cdot10^{-10}$ Torr during epitaxy\cite{kuntsevich2016,oveshnikov}. Ternary Sr$_x$Bi$_{2-x}$Se$_3$ layers were deposited using standard effusion cells for high purity elemental Se, Sr and binary Bi$_2$Se$_3$ compound. Atomic/molecular fluxes were calibrated using Bayard-Alpert ion gauge that swings into the substrate position. Stability of the beam equivalent pressure (BEP) for each cell was controlled just before starting the deposition and immediately after the growth finish. The cells temperatures for Se ($130^\circ$C) and Bi$_2$Se$_3$ ($495^\circ$C) were held constant, that resulted in Se-rich conditions with a BEP flux ratio Se/Bi$_2$Se$_3$ of 2:1 and provided layer growth rate of ~ 0.25 nm/min. Thin films with different Sr content were obtained by varying the evaporation temperature of the Sr effusion cell.
Sr concentration $x$ in the grown films was increased from $0.003$ to $0.352$ when heating Sr cell from 260 to $380^\circ$C. When growing Bi$_2$Se$_3$ films from elemental Bi and Se sources \cite{vicinalSi111,OhSi111,OhAl2O3,AndoAl2O3,brasBaF2} it is inevitable to employ a two-step deposition method due to a weak interaction of Bi and Se ad-atoms with the substrate, with the first step temperature being below 150 $^\circ$C. Beam flux composition, generated from effusion cell loaded with binary Bi$_2$Se$_3$ compound \cite{kuntsevich2016,oveshnikov,JMMM331}, consists of Bi-Se molecular species\cite{SpringholzBaF2}, and thin film condensation occurs at the elevated temperatures. So, time consuming low temperature stage becomes unnecessary. We found that ``ramped up'' approach, described in the next paragraph, provides a higher quality ternary films as
compared to straight 300$^\circ$C single step deposition.

Growth of the film was started at $260^\circ$C with opening Se and Bi$_2$Se$_3$ cell shutters. Following the 4 minute deposition of the first quintuple layer (QL) of binary Bi$_2$Se$_3$, substrate temperature was ramped up to $300^\circ$C in the next 4 minutes without growth interruption. Then Sr cell shutter was opened for the deposition of the body of the ternary film on top of the 2QL thick buffer layer. Thin films with different Sr composition $x$ and thicknesses of 20-50 nm were obtained by the described approach. Immediately after the growth, the film on the substrate was cooled down to the room temperature and, in order to protect the film surface from ambient atmosphere, an amorphous 30~nm thick Se cap layer was deposited.
Concentration $x$ of Sr in the Sr$_x$Bi$_{2-x}$Se$_3$ layers was  determined from the fluxes, generated by Sr and Bi$_2$Se$_3$ cells during growth. For Sr cell temperatures less than 360~$^\circ$C, BEP signal drops below the ion gauge sensitivity. So, measurements were done for temperature range from 360~$^\circ$C to 550~$^\circ$C, and obtained BEP dependence on cell temperature was approximated for the specified range from 360~$^\circ$C down to 260~$^\circ$C. In order to get absolute calibration for the flux rate of Sr, a set of epitaxial SrSe (FCC) layers on (111) BaF$_2$ was grown at two growth rates: 0.3 and 1.0 nm/min. To convert BEP signals into atomic fluxes, Sr flux rate was determined from SrSe reference film thicknesses measured at (0 0 2) XRD reflection oscillations and film deposition time.
 A series of Sr$_x$Bi$_2$Se$_3$ bulk crystals with nominal Sr content of $x$=0, 0.01, 0.037 and 0.15 were prepared using modified Bridgeman method~\cite{Aleshchenko,KuntsevichNematic}. To compare properties of thin films and bulk crystals, within this paper we define nominal composition  $x$ as a molar fraction of dopant $x$ times 100\%(below we will give nominal $x$ in percentage).

\subsection{\label{xrdtech}X-ray measurements}
The X-ray diffraction(XRD) and X-ray reflection(XRR) measurements
were carried out on Panalytical MRD Extended diffractometer with a
hybrid monochromator, that is a combination of a parabolic mirror
and a single crystal 2$\times$Ge($2\,2\,0$) monochromator. We used
triple crystal-analyzer 3$\times$Ge($2\,2\,0$) to get high resolution ($2\theta-\omega$)-scanning curves for lattice parameter determinations.
 Thickness of the films was obtained from ($0\,0\,6$)  Bragg peak diffraction fringes and/or from X-ray reflection (XRR) spectra.

\begin{table*}[ht!]
\centering
 \begin{tabular}{c c  c  c c c  c c c c}
 \hline\hline
 Sample & X$_{Sr}$($\%$) & d (nm) & Parameters $c_{0\,0\,6}$,$c_{0\,1\,5}$,$a_{0\,1\,5}$& RRR &n (cm$^{-2}$) &$\mu$ (cm$^2$/Vs)  & $\bigtriangleup\omega_{0\,0\,6}$ & $\bigtriangleup\omega_{0\,0\,15}$& $\bigtriangleup c/c$\\
 \hline
 771 & 0 & 25.4 &$c_{0\,0\,6}$=28.710 $c_{0\,0\,15}$=28.724 $a_{0\,1\,5}$=4.1305 &1.53& $3.1\cdot10^{13}$&1400 &0.158&0.199 &0.049 \\
  763 & 0.3 & 25.9 &$c_{0\,0\,6}$=28.706 $c_{0\,0\,15}$=28.722 $a_{0\,1\,5}$=4.129 &1.51& $1.5\cdot10^{13}$&2989 &0.165 & 0.229 &0.056\\
 762 & 0.8 & 30.7 &$c_{0\,0\,6}$=28.754 $c_{0\,0\,15}$=28.764 $a_{0\,1\,5}$=4.125 &1.48& $1.0\cdot10^{13}$&800 &0.214 & 0.279 &0.035\\
 757 & 1.8 &  23.7 &$c_{0\,0\,6}$=28.811 $c_{0\,0\,15}$=28.822 $a_{0\,1\,5}$=4.125 &1.64& $4.7\cdot10^{13}$&681 &0.171 & 0.239 &0.038\\
  747 & 3.6 &  30 &$c_{0\,0\,6}$=28.805 $c_{0\,0\,15}$=28.824 $a_{0\,1\,5}$=4.126 &1.50& $9.7\cdot10^{13}$&389 &0.190 & 0.41 &0.066\\
 746 & 10.5 &  23 &$c_{0\,0\,6}$=28.956 $c_{0\,0\,15}$=28.936 $a_{0\,1\,5}$=4.117 &1.43& $8.9\cdot10^{13}$&142 &0.320 & 0.46 &-0.069\\
   778 & 15.7 & 29.2 &$c_{0\,0\,6}$=29.073 $c_{0\,0\,15}$=28.879 $a_{0\,1\,5}$=4.115 &1.4& $2\cdot10^{14}$&120 &0.658 & 0.868 &-0.67\\
  745 & 25.8 & 35.6 &$c_{0\,0\,6}$=29.487 $c_{0\,0\,15}$=28.979 &-& -&- &-&-&-1.75\\
  748 & 35.2 & 25 &$c_{0\,0\,6}^1$=28.714 $c_{0\,0\,6}^4$=29.927  &1.03& $1.5\cdot10^{14}$&37 &- &-&- \\
 \hline\hline
 \end{tabular}
 \caption{Summary of sample parameters. The amount of Sr in Bi$_2$Se$_3$ films X$_{Sr}$ was calculated from the flux ratios. Thickness $d$ and parameters $a,c$ were determined from XRD. Carrier density($n$) and mobility ($\mu$) were obtained from low-field Hall measurements (at $T=2K$). Residual-resistivity ratio(RRR) was calculated as room temperature resistivity($R_{T=300K}$) ratio to the low temperature resistivity minimum.}
\label{table:Table1}
\end{table*}

\begin{figure}
\includegraphics[width=8.5 cm]{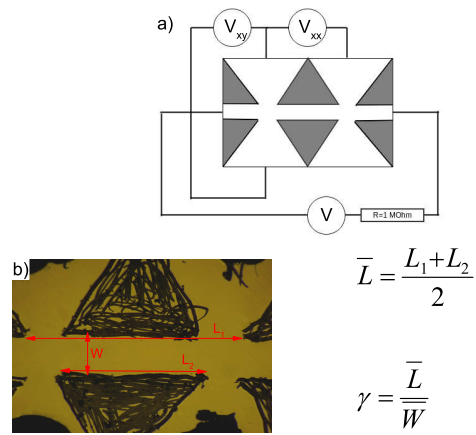}% Here is how to import EPS art
\caption{\label{figure:Fig1} a) Schematics of electrical connections used for transport measurements. b) Image of sample. The arrows show the geometrical definition of width and length.}
\end{figure}

\subsection{\label{afmtech}Atomic force microscopy measurements}
The morphology of the representative films was explored using the atomic force microscopy (AFM). For these measurements selenium capping layer was removed by annealing a sample in a vacuum at 200$^\circ$C for 10 minutes.  Single crystal was freshly cleaved for morphological studied. The measurements were performed using NT-MDT Solver 47 Pro system in tapping mode at ambient conditions.

\subsection{\label{transptech}Transport measurements}
 Hall-bar film geometries for transport measurements were defined by scratching the films with needle, similarly to Ref.\cite{kuntsevich2016}. Samples  were mounted on the holder and contact wires were attached with silver paint (contact resistance was typically below several hundreds Ohms).  Geometrical factor ${l}/{w}$ for every sample was evaluated from the camera image of the Hall bar, as shown in Fig. ~\ref{figure:Fig1}b. Low-temperature magnetotransport measurements were performed using a standard lock-in technique at frequencies $13-80$ Hz and measurement current 1 $\mu$A to ensure the absence of overheating at the lowest temperatures.  All measurements were performed in the temperature range 1.6 K-300 K using Cryogenics dry CFMS-16 system.  Perpendicular magnetic field was swept  (typically at 2,4,8,16,32 K) from positive to negative value(typically 1T). In order to compensate contact misalignment, the  magnetoresistance (Hall resistance) data were symmetrized (antisymmetrized). Using the $\rho_{xx}(B)$ and $\rho_{xy}(B)$ dependencies we determined  the carrier density, Hall mobility and investigated weak antilocalization. The main structural and transport parameters are summarized in Table~\ref{table:Table1} and discussed below in Results and Discussion sections.

\section{\label{expresults}Experimental results}

\subsection{\label{structmorphresults}Structural and morphological results}
\begin{figure}
\includegraphics[width=8.5 cm]{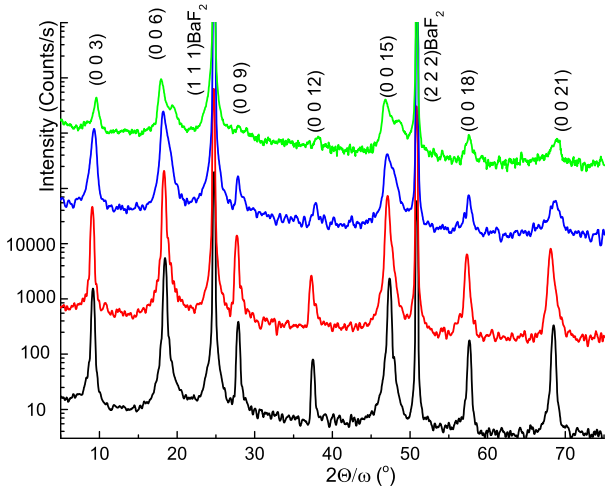}
\caption{X-ray diffraction scans for epitaxial films with different Sr content x (in \%, from bottom to top), specified by different colors: 0\% - black, 1.8\% - red,  14\% - blue, and 25.8\% - green.}
\label{Fig9}
\end{figure}

 In Fig.~\ref{Fig9} diffractogramms ($2\theta-\omega$)-scans of four representative samples with different Sr composition(listed in Table~\ref{table:Table1}) are shown. Besides the intensive ($1\,1\,1$), ($2\,2\,2$) reflections from BaF$_2$ substrate (partially cut for better visibility of the film signal)
 a series of ($0\,0, l$) reflections from the film are clearly seen. They evidence for the  growth of highly oriented layers with basal plane ($0\,0\,1$) parallel to the  BaF$_2$ substrate ($1\,1\,1$) cleavage plane. As Sr content $x$ rises from $x=0\%$ to $x=14\%$,  the ($0\,0\,l$) peaks  get wider, indicating the increase
  of structural disorder (see columns  $\bigtriangleup\omega_{0\,0\,6}$, $\bigtriangleup\omega_{0\,0\,15}$ of Table~\ref{table:Table1}).
  At the same time,  the reflections with $l\leq$15  shift to smaller diffraction angles.   For $x \approx$25.8\% in addition to broadening, the ($0\,0\,6$) and ($0\,0\,15$) reflections split, while the intensities of ($0\,0\,9$) and ($0\,0\,12$) ones get suppressed. Thus, $x$=25.8\% sample already  consists of crystallites with different Sr composition, and film becomes inhomogeneous.

\begin{figure}[h!]
\begin{minipage}{3.2in}
\includegraphics[width=8.5 cm]{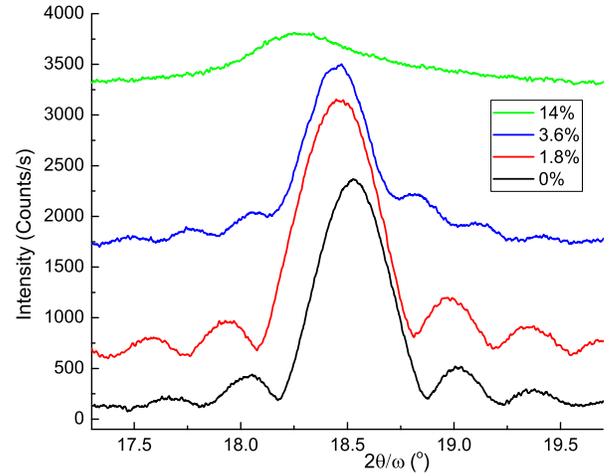}
\caption{2$\theta-\omega$ XRD scans on (0 0 6) reflex for films with different $x$ content increasing from bottom to top, indicated in the panel.}
\label{Fig10}
\end{minipage}
\end{figure}

High resolution ($2\theta-\omega$) scans of ($0\,0\,6$) reflection are shown in Fig.~\ref{Fig10}. The peak position is clearly seen to move to smaller angles with $x$, indicating that $c$-lattice parameter substantially grows as doping level increases from 0\% to 14\%. Coming back to Fig.~\ref{Fig9} one can see that for $x>13$\%, the ($0\,0\,18$) and ($0\,0\,21$) reflections are shifted to the opposite direction (high $\theta$ values). This observation might be due to decrease of the XRD intensity from the defective highly Sr-doped fragments of the films.
%Probably, this fact is due to decreasing the XRD intensity from the defective highly Sr-doped fragments of the films.

For low Sr doping level(below 4\%) we observe intensity fringes
near the central($0\,0\,6$) peak (see Fig.~\ref{Fig10}). Their period
straightforwardly gives the film thickness\cite{SCC}, $L=\frac{\lambda}{2 \times (\sin{\omega_2}-\sin{\omega_1})}$,
where $\omega_1$ and $\omega_2$ are neighboring maxima positions,and $\lambda$ is a X-ray wavelength. Intensity of the main reflection and its satellites decreases with $x$. Satellites are fully smeared by structural disorder for doping levels above 4\%.

\begin{figure}[h!]
\begin{minipage}{3.2in}
\includegraphics[width=8.5 cm]{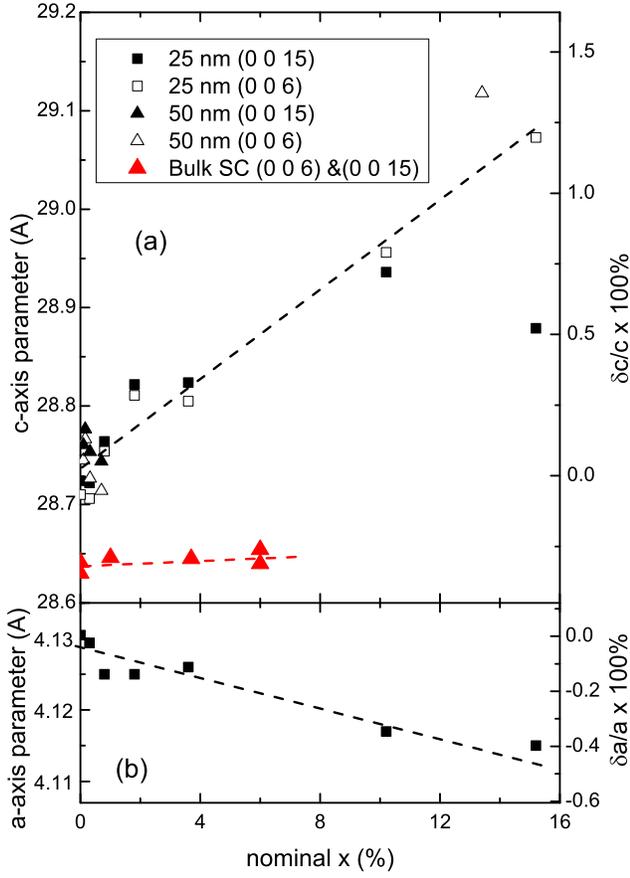}
\caption{{{(a) Summary of the $c$-axis lattice parameter value}} on
($0\,0\,6$) and ($0\,0\,15$) XRD reflections as a function of $x$ for 25~nm and 50~nm films. Black filled symbols correspond to ($0\,0\,15$) reflection, empty ones to ($0\,0\,6$) reflection. Squares correspond to $\sim 25$~nm thick films, and  triangles correspond to $\sim 50$~nm thick films. Triangles show c-parameter for single crystals.
Straight lines show approximations of $c(x)$ dependencies for  bulk crystals and thin films with the slopes 0.2 pm/\% and $\approx$ 2.05 pm/\% respectively (points with $x>12\%$ are not used for approximation). (b) $a$-axis parameter dependence on x for 25 nm thick films. Straight line is a linear interpolation. Right axes in panels (a) and (b) show the relative variation of the lattice parameter.} \label{Fig200}
\end{minipage}
\end{figure}

Fig.~\ref{Fig200} a shows the doping level dependence of the $c$-lattice parameter determined from both ($0\,0\,6$) and ($0\,0\,15$) reflections.
 There is a very small difference between  $c_{0\,0\,6}$ and $c_{0\,0\,15}$  for $x$ below $12$\%. Above $x=12$\% the $c_{0\,0\,6}$ and $c_{0\,0\,15}$
 values start to differ dramatically, that indicates that above 12\% crystal structure becomes more defective. The monotonic increase
 of $c$-parameter by approximately 1$\%$ with $x$ (for $x\approx12$\%) can be well interpolated by linear approximation with the slope
 $\frac{dc}{dx}=$2.05pm/\% ( black dashed line Fig.~\ref{Fig200}a). Note, that for the bulk crystals (red dashed line) the slope is 10 times smaller, $\frac{dc}{dx}=$0.2pm/\%,  that is in beautiful agreement with results of Ref.\cite{joaf}. The absolute value of $c$-lattice parameter for Sr-doped crystals 28.63 A $\div$28.65 A is also in good agreement with previous studies (28.65 A in Ref.\cite{deVisser}; 28.64 A in Ref. \cite{shruti}; 28.664 A in Ref.\cite{smylie}).
  At the same time, in our films $a$- lattice parameter decreases with $x$ (see Fig.~\ref{Fig200} b).  The combination of strong $c$-axis expansion and $a$-axis contraction with $x$ indicates that Sr atoms predominantly substitute Bi atoms or occupy interstitial sites within quintuple-layer (QL), rather than intercalate into  Van-der Waals(VdW) gap (see section \ref{discus}).

\begin{figure}[h!]
\begin{minipage}{3.2in}
\includegraphics[width=8.5 cm]{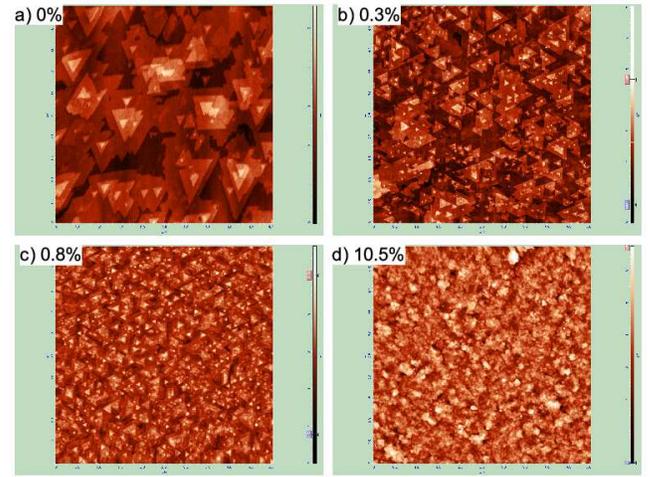}
\caption{Atomic force microscope patterns (5$\times$5 $\mu$m$^2$) for 25~nm thick films with different doping level:
a)Undoped Bi$_2$Se$_3$. b) $x$=0.3\%. c) $x$=0.8\%. d) $x$=10.5\%.}
\label{Fig3}
\end{minipage}
\end{figure}

With that strong  variation of $c$-lattice parameter accompanied by structural degradation we may expect a pronounced evolution of thin film  morphology with composition. Fig.~\ref{Fig3}  shows the AFM ($5\times5$~$\mu$m$^2$) scans of four representative 25~nm thick films with  various Sr doping level. Undopped film demonstrates a
rather big triangular domains with flat terraces and linear dimensions above 1~$\mu$m (RMS=0.64~nm). These domains have predominantly  the same orientation(see Fig.~\ref{Fig3}a) reflecting a rather small twinning-level.  The size of domains tends to decrease with $x$. In addition, concentration of twin domains substantially increases,  as seen from Fig.~\ref{Fig3}~b,c. An average domain size is 280 nm (RMS=0.61), and drops to 190 nm (RMS=0.97)  for $x$=0.3\% and 0.8\%, respectively.   For more heavily doped (10.5$\%$) film, the in-plane  triangular domain structure is not AFM resolved(see Fig.~\ref{Fig3}d). A rather similar morphological transformations were observed on doping
Bi$_2$Se$_3$ with Ca \cite{ptype}.

\begin{figure}[h!]
\begin{minipage}{3.2in}
\includegraphics[width=8.5 cm]{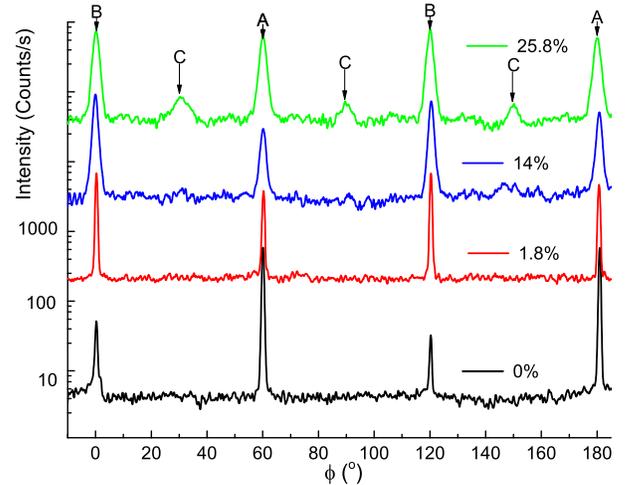}
\caption{{\bf{$\varphi$ scan}} about of the [$0\,0\,1$] axis on (1
0.10) reflection for films with different $x$ (0\%,1.8\%, 14\%,
and 25.8\% from bottom to top).  Angular positions of primary (A) and secondary (B) twin domains are indicated by arrows. Amount of  60$^\circ$ twins (B)
increases with $x$. For $x$= 25.8\%, 30$^\circ$ twins (C) emerge.}
\label{Fig11}
\end{minipage}
\end{figure}

Evolution of the film domain structure with $x$ is clearly confirmed by XRD $\varphi$ scans about the [$0\,0\,1$] axis on asymmetric ($1\,0\,10$) reflection. Instead of 120$^\circ$ periodic reflections, as expected for  rhombohedral structure (Fig.~\ref{Fig11}), we see peaks repeated  every 60$^\circ$ (A and B). Ratio between amplitudes of two A and B sets of peaks reflects twinning degree. In the undoped Bi$_2$Se$_3$ film, the intensity ratio of peaks for different 60$^\circ$-twin domains is 6:1. In $x\approx 1.8\%$ film this ratio drops to $1.7:1$ and further tends to 1:1 as $x$ increases. All reflections in   $\varphi$-scan curves widen with $x$, indicating a decrease in the coherent X-ray scattering regions along the diffraction direction. This fact is in a good agreement with the decrease in the dimensions of the flat terraces in the basal plane, visible by atomic force microscopy. In addition to 60$^\circ$ twins, the 30$^\circ$ rotational domains (reflections C in Fig.~\ref{Fig11}) also appear with the increase of Sr concentration to 25.8\%.

\subsection{\label{restransp}Transport measurements}
\begin{figure}[h!]

\begin{minipage}{3.2in}

\includegraphics[width=8.5 cm]{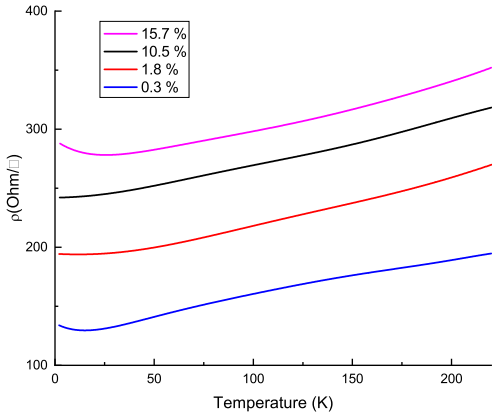}

\caption{Resistivity vs. temperature for four representative Sr doped Bi$_2$Se$_3$ films ($x$=0.3\%, 1.8\%, 10.5\%, and 15.7\% from bottom to top). All $\rho(T)$ dependencies exhibit metallic behavior with RRR$\approx1.5$.}
\label{RT}
\end{minipage}
\end{figure}

Transport properties also show up systematic dependencies on $x$. The resistance per square ($\rho$) vs temperature for the representative  films is shown in Fig.~\ref{RT}. All films have metallic type resistivity ($\frac{d\rho}{dT}>0$) in wide range of temperatures. The value of the resistivity per square tends to enhance progressively with Sr doping level for all temperatures. The residual-resistance ratios(RRR), defined as ratios between the resistance at 300~K and minimal resistance at low $T$, are summarized in Table~\ref{table:Table1}.
RRR is about $\sim 1.5$ for most of the 25~nm thick films. This fact indicates the similar scattering mechanisms in all samples. We observe the minimum  of $\rho$ and its low-$T$ upturn, caused by e-e interaction, similarly to numerous observations on undoped Bi$_2$Se$_3$ films\cite{kuntsevich2016}. Note that, we do not observe any signs of superconductivity down to 1.6K.

\begin{figure}[h!]
\begin{minipage}{3.2in}
\includegraphics[width=8.5 cm]{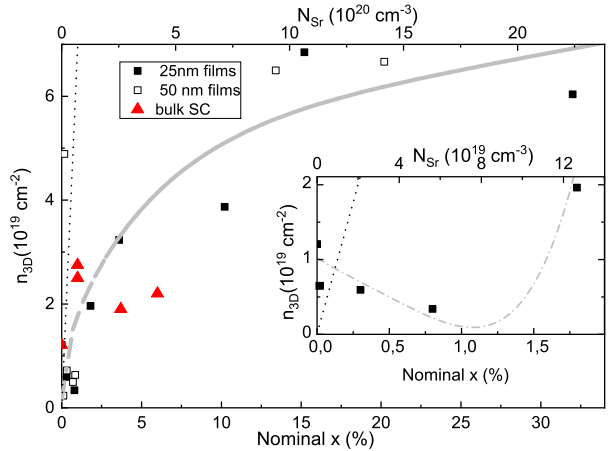}
\caption{Hall carrier density (at $T=2K$) as a function of doping level for 25~nm thick films (black squares). Red triangles correspond to Sr$_x$Bi$_2$Se$_3$ crystals. Top axis shows the corresponding bulk density of Sr atoms. The inset shows a zoom-in of the low-$x$ region. Dash-dotted lines are interpolated $n(x)$ dependencies. Dotted straight lines correspond to Sr doping efficiency $n_{3D}/N_{Sr}=1$.}
\label{Fig421}
\end{minipage}
\end{figure}

\begin{figure}[h!]
\begin{minipage}{3.2in}
\includegraphics[width=8.5 cm]{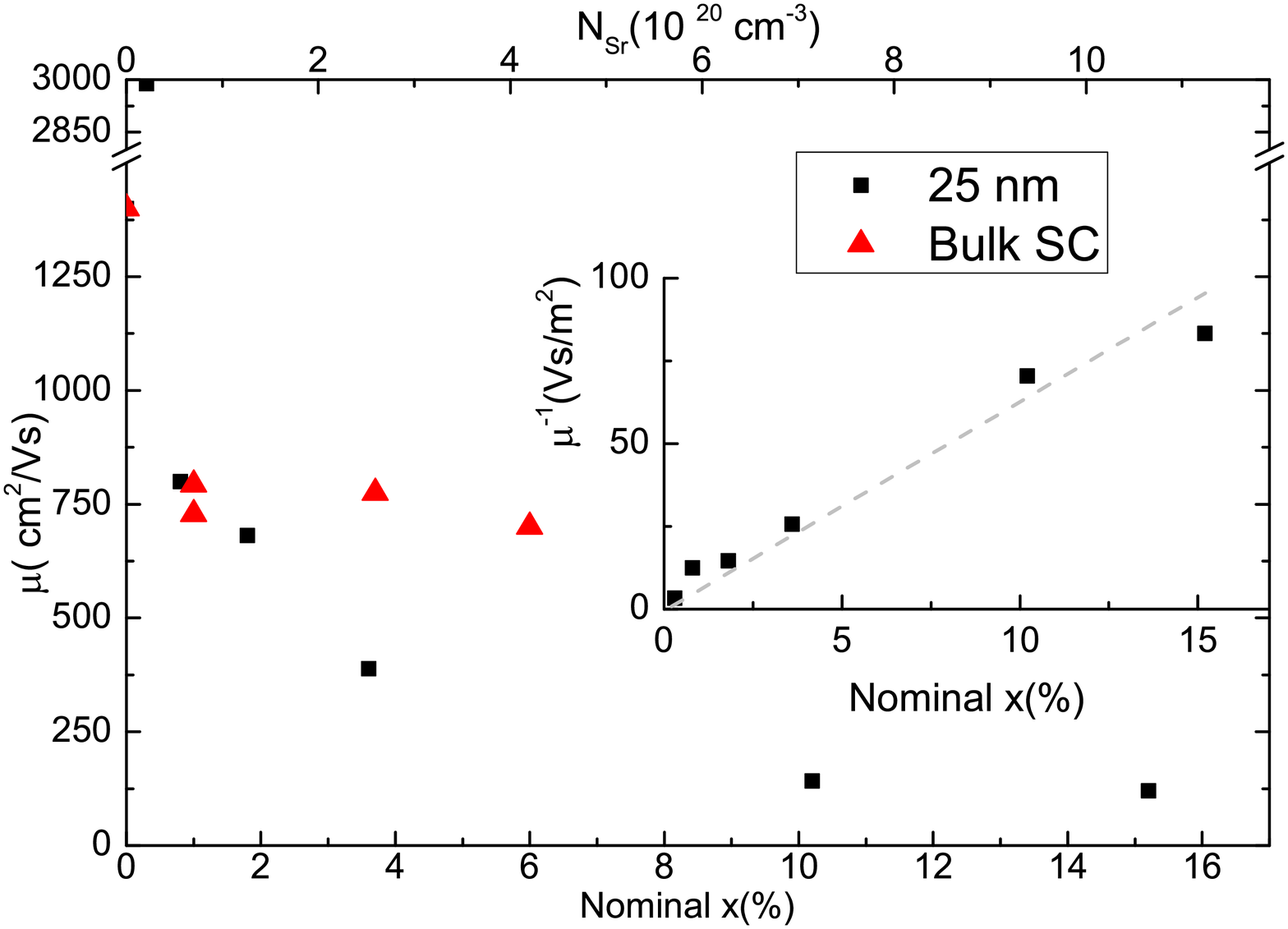}
\caption{Hall carrier mobility (at $T=2K$) as a function of doping level for 25~nm thick films(black squares). Red triangles correspond to Sr$_x$Bi$_2$Se$_3$ crystals.  The insert shows approximately linear $x$-dependence of the inverse mobility for 25-nm thin films.}
\label{Fig4}
\end{minipage}
\end{figure}

As obtained from the sign of the Hall effect, all films are n-type. In Fig.~\ref{Fig421}, and Fig.~\ref{Fig4} the boxes show carrier density and mobility as a functions of $x$, respectively.  Importantly, after some low-x drop (see insert to Fig. \ref{Fig421}) carrier density generally increases with $x$, whereas mobility decreases for all $x$.  The increase in density is much weaker than the decrease in mobility, so the resistivity (inverse product of carrier mobility and density) increases with $x$ (Fig.~\ref{RT}).
 It proves that (i) not each Sr atom act as a donor (moreover, the larger value of $x$ the weaker the doping effect, carrier density saturates); (ii) At the same time number of scattering centers (inverse mobility, see insert to Fig.\ref{Fig4}) grows roughly linearly with $x$. We believe, that Sr atoms in the lattice act at least in two ways simultaneously: in one position (probably substitution of Bi) they act as acceptors, whereas in the other position ( probably interstitial) they act as donors. It is also possible that Sr promotes formation of Se vacancies, also acting as donors. Anyway the doping efficiency of Sr is very small (compare $n(x)$ and dotted straight lines in Fig.~\ref{Fig421}), Sr atoms mostly act as a scattering centers and mobility is inversely proportional to $x$.

\subsection{\label{wal}Quantum transport}
For low temperatures, a pronounced $B=0$ dip in magneto-resistance  was observed for all studied films. Similarly to previous studies on thin films of Bi-chalchogonies\cite{cufilms, cufilms3, kuntsevich2016}, we attribute it to weak antilocalization. We fitted magneto-conductivity with Hikami-Larkin-Nagaoka(HLN)\cite{HLN} formula: $$\sigma(B)-\sigma(0)=-\frac{\alpha e^2}{2\pi^2 \hbar}[\psi(\frac{1}{2}+\frac{\hbar}{4eBl_\varphi^2})-\ln(\frac{\hbar}{4eBl_\varphi^2}))]$$ Here, $\alpha$ is an adjustable prefactor; $e$ ,$\hbar$ stay for electron charge and Plank constant, respectively; $L_\varphi$ is adjustable phase coherence length, $\psi$ is the digamma function.

\begin{figure}[h!]
\begin{minipage}{3.2in}
\includegraphics[width=8.5 cm]{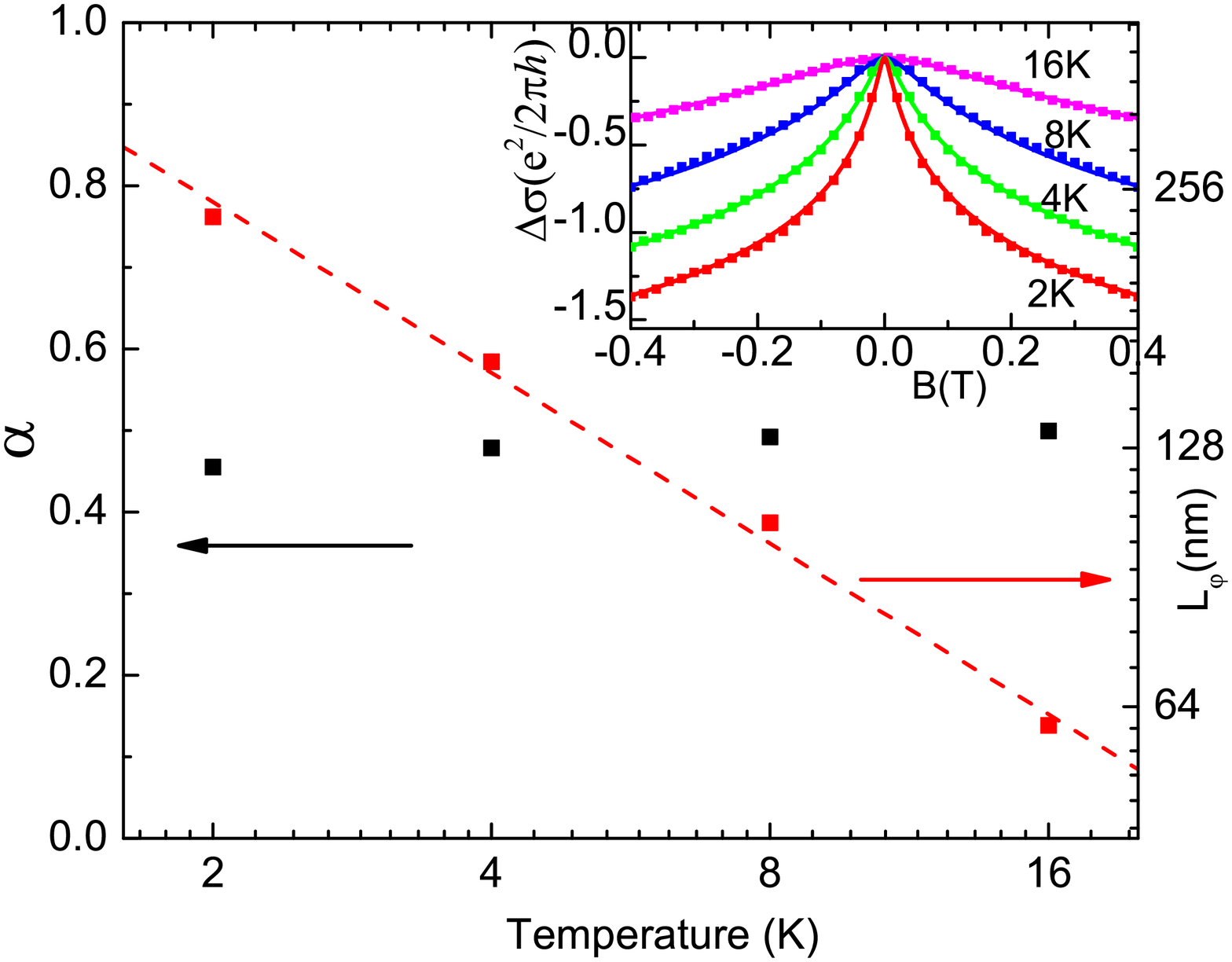}
\caption{The insert shows the  weak antilocalization magnetoconductance at  different temperatures (symbols) for $x$=10.5\% film (746) and their fits with HLN formula. Main panel shows the fitting parameter $\alpha$ (left axis, black squares) and phase breaking time $l_\varphi$ (red squares) as a function of temperature. The dashed line is $\propto T^{-0.5}$ fit.}
\label{figure:Wallfigure}
\end{minipage}
\end{figure}

The inset in the Fig.~\ref{figure:Wallfigure}  shows a set of low-field magnetoconductivity fits for various temperatures for representative 25 nm thick film($x=10.5\%$). The fitting parameter $\alpha \approx 0.5$ (indicated in the inset) does not depend on temperature, similarly to the others\cite{cufilms, kuntsevich2016}. The coherence length decreases with temperature as $\sim T^{-0.5}$ (see red dashed line in Fig.~\ref{figure:Wallfigure}), indicating the electron-electron scattering mechanism of dephasing.

The other Sr-doped films demonstrate similar values of $\alpha$ and $L_\varphi(T)$ dependencies. This observations are consistent with numerous reports on non-doped films of Bi chalcogenides \cite{kuntsevich2016,Wanga,Onose}. Thus, our studies indicate that Sr impurities do not affect the quantum transport.

\section{\label{comparison}Comparison of epitaxial films and bulk superconducting single crystals }

\begin{figure}[h!]
\begin{minipage}{3.2in}
\includegraphics[width=8.5 cm]{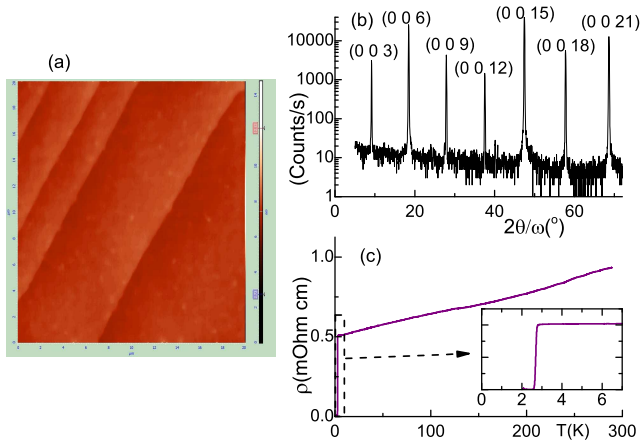}
\caption{(a)Representative morphology of the freshly cleaved superconducting Sr$_{0.06}$Bi$_2$Se$_3$ single crystal with nominal Sr content $x=0.15$ (20$\times$20 $\mu$m$^2$ scan) with 1~nm QL steps easily distinguishable; (b) $2\theta/\omega$ XRD scan; (c) $\rho(T)$-dependence with the zoom in of the superconductive transition in the inset.}
\label{figure:singcryst}
\end{minipage}
\end{figure}

Absence of superconductivity in thin films motivated us to perform a detailed comparison with SC single crystals.
 The single crystals have perfect, almost single domain structure (as seen from AFM scan in Fig.~\ref{figure:singcryst}a), perfect crystal quality (as seen from XRD measurements Fig.~\ref{figure:singcryst}b), demonstrate superconductivity and RRR$\sim 1.8$ (Fig.~\ref{figure:singcryst}c).

For all SC single crystals typical Hall density is about $2\times10^{19}$cm$^{-3}$ and the critical temperature  $T_c$ is in the range from 2.4~K to 3~K.  These properties are highly reproducible in a number of different labs~\cite{KuntsevichNematic,deVisser,shruti,joaf}.

{The remarkable feature of the crystals, mentioned in Ref.~\cite{joaf}}, is that the Sr real content saturates around 6\%, that is nearly independent of nominal composition $x$, for $x$ above 10\%.

Sr-doped Bi$_2$Se$_3$ always remains n-type~\cite{joaf,shruti,KuntsevichNematic,pso}, despite the fact that divalent Sr replacing trivalent Bi should act as a strong acceptor. These strange issue is related to a rather complex tetradymite lattice structure. Dopants may be distributed over a large number of electrically inequivalent incorporation sites within unit cell consisting from three QLs separated by Van-der Waals gaps.

In the thin films we can tune both composition and Hall density beyond $x$=6\% and $n=2\times10^{19}~cm^{-3}$, respectively. Yet, we do not observe superconductivity.

Free carrier concentration of $n\approx2\times10^{19}~cm^{-3}$ is achieved already at $x$ $\approx 2\%$, in both, thin films as well as bulk crystals. However, at higher Sr content in bulk crystals $n$ is stable with $x$, while it reaches n$\approx 7\times10^{19}~cm^{-3}$ in thin films and saturates at that level for $x$ above 12\% (compare boxes and triangles in Fig.~\ref{Fig421}). These results suggest various mechanisms of doping in crystals and thin films.

It is not surprising therefore that mobilities (see Fig.\ref{Fig4}) also display different behavior. In thin films mobility dramatically drops, while in the same range of $x$ between 1\% and 6\% scattering of carriers in bulk crystals is nearly composition insensitive.

Interestingly, a much more pronounced distinction shows up in the crystal structure (see Fig.~\ref{Fig200}): as $x$ increases, $c$-lattice parameter in epitaxial films grows an order of magnitude faster than in bulk crystals (see triangles versus boxes in Fig.~\ref{Fig200}). This is a direct indication that Sr takes different positions in the lattice of the films and bulk crystals.

Another essential difference between SC crystals and thin films is the domain structure. Indeed, while the crystals consist of slightly misaligned, hundreds-of-microns sized blocks (see morphology in Fig\ref{figure:singcryst}a)\cite{KuntsevichNematic, smylie}, Sr-doped films are composed of submicron-sized triangular twin domains with two opposite orientations. These domains are evidenced by the AFM pictures in Fig.\ref{Fig3} and XRD Phi-scans (Fig.~\ref{Fig11}). Unless special care is taken~\cite{tarakina,guo}, even binary undoped Bi$_2$Se$_3$ films on lattice matched InP substrates are heavily twinned.  In our case, concentration of twin domains tends to increase with Sr content. Twinning means that the anti-phase domain boundaries interpenetrate the whole body of the film, thus promoting orders of magnitude faster diffusion of foreign atoms and governing redistribution of Sr atoms in the lattice.
It is important therefore to understand whether domain boundaries and elevated c-lattice constant (Sr content and placement) in thin films are interrelated. Observation of novel properties (and superconductivity in the optimistic scenario) for single-domain Sr-doped films would be crucial experiment, that clarifies the role of grain boundaries.  The growth of single-domain Sr-doped Bi$_2$Se$_3$ films however calls for sophisticated substrate interface engineering and has not been performed so far.

\section{\label{discus}Discussion}

 Crystal structure of Bi$_2$Se$_3$ is constructed from QLs (five layer Se1-Bi-Se2-Bi-Se1 sandwiches)  van der Waals bonded to the neighboring QLs. This layered nature allows the dopants to occupy not only substitutional or interstitial sites in the host material, but also enter the Van-der Waals gap between the QLs. So, Sr atoms may reside on different positions.

First, we discuss formation of substitution defects Bi$_{\rm sub}$ in our thin films. With increasing Sr doping level, as shown in Fig.\ref{Fig200}a,b, an expansion in the $c$-axis is accompanied by the $a$-axis contraction. We argue, that the expansion along the $c$-axis was most likely not a result of Sr-intercalation. Indeed, our experimental data indicate that when Sr content reaches $x\sim10$\% the $c$-axis relative increase taken at (0 0 15) peak $\delta c/c = +0.7\%$ , while the $a$-axis decrease calculated for (0 1 5) reflection $\delta a/a = -0.33\%$. So, with obtained relation $\delta c/c \approx -2\delta a/a$, volume of the unit cell, given in hexagonal notation by $V=a^2\cdot c\cdot sin(60^\circ)$, remained constant. This is consistent with Bi substitution by Sr atoms, while intercalation of Sr into the Van-der Waals gap is unlikely. Another manifestation of substitution Sr$_{Bi^{-1}}$ defect is given in the inset of Fig.\ref{Fig421}. In the low Sr content region ($x$ up to $\sim 1$\%), free carrier concentration $n$ drops nearly four times indicating Sr$^{2+}$ substitution into Bi$^{3+}$ sites. However, the decrease in $n$ with Sr concentration is much lower than expected if all Sr atoms behave as acceptors. Our structural, morphological and mobility data indicate that introduction of Sr also adds structural defects, such as Se vacancies, anti-phase domain boundaries or even Sr interstitial impurities, for example. These structural defects may add carriers to the conduction band while Sr$_{Bi^{-1}}$ act as acceptors. So, different structural defects induced by Sr incorporation into the host crystal lattice compensate acceptors produced by Sr substitution into Bi sites. For $x > 1$\% n-type doping prevails, and interplay between dominating donors and Bi$_{sub}$  acceptors saturates at the level $\sim 6\times 10^{19}$~cm$^{-3}$ when material becomes inhomogeneous.

Next, we discuss a possibility of Sr$^{+1}$ $_i$ defects in our films. Covalently bonded inside QL arrangement of Bi and Se atoms provides a plenty of space for Sr$^{2+}$ (ionic radius 118 pm) to reside between Bi-Se1 or Bi-Se2 planes, with inter-atomic distances of 284~pm and 304~pm \cite{lind}, respectively. So, interstitial defects are rather expected. Interstitial incorporation of Mn and Fe atoms was observed in MBE grown Bi$_2$Se$_3$ films by means of X-ray absorption fine structure (XAFS), even at low doping concentrations\cite{Figueroa}.
First-principles calculations of the formation energies corresponding to different Sr doping locations in Bi$_2$Se$_3$ lattice predict interesting situation \cite{pso}: At low Sr concentration the most stable doping sites are Bi substitution and Van- der Waals intercalation, while Bi-Se2 interstitial position is energetically less stable. As Sr concentration rises twofold, formation energy of Bi-Se2 interstitial position increases threefold and catch up with those of Bi$_{sub}$ and Van- der Waals intercalation doping. So, at higher doping level Bi-Se2 interstitial position becomes more stable. This is consistent with the observed n-type doping behavior in our thin films for Sr content above $x=$1\%. Absolute values of the calculated formation energies are $\sim-41$ meV for Bi$_{sub}$ and Bi-Se2 sites, and $\sim~-49$ meV for Van-der Waals position\cite{pso}. That small difference could be of no consequence for bulk single crystals, and may play a decisive role for thin films.

Apparently, the growth conditions in MBE and Bridgeman method are absolutely different: thin films are epitaxially assembled in vacuum at $\approx$570~K, with an excess flux of Se. This extra flux is unavoidable, and it is aimed to maintain stoichiometry of the growing Bi$_2$Se$_3$ film even in the presence of additional flux of Sr atoms. Growth of the bulk crystals starts by melting of encapsulated constituent elements Sr:Bi:Se with a molar ratio $x$:2:3 at a temperature 1120~K and ends at 900~K. So, the system stoichiometry is shifted towards the metal excess, and temperature is twice as large as for thin films. These two factors may provide much more probability for Sr dopant atoms to attain the most energetically favorable positions in the lattice. For thin films, at temperatures above 570~K thermal etching already starts to destroy the layer, making higher temperatures impossible. So, the dopant adatoms at the growing surface meet excess Se and may have not enough energy to occupy a ``proper'' superconducting-relative sites.
Another reason why thin films do not attain superconductivity, could be connected with energetic stability of the ``proper'' sites. Indeed, while bulk crystals are as a rule water quenched, thin films are gradually cooled down with substrate holder. Authors in \cite{pso} observed, that low doped (5\%) crystals chilled out from 620~$^\circ$C to room temperature in the furnace turned into not superconducting crystals, contrary to the quenched samples.

It was suggested initially that superconductivity in bulk Sr-doped Bi$_2$Se$_3$ crystals is achieved through intercalation. In other words, observed slight increase in the $c$-axis constant~\cite{shruti,joaf}  implies that Sr dwells in the Van-der Waals gap. Later, from  scanning tunneling spectroscopy experiments~\cite{Ncomms2017}  two possible locations for Sr were derived: either intercalated in Van der Waals gap or placed inside quintuple layer.

In addition, from atom-by-atom elemental analysis based on EDX mapping it was concluded, that in addition to
substitution defect Sr$_{Bi}^{-1}$, dopant atoms may form Sr$_i^{+1}$ various interstitial defects. Microstructural TEM investigations of Sr-doped Bi$_2$Se$_3$ thin films~\cite{chinawork} revealed, that distance between Bi atomic
sheets inside QL was compressed, while separation between the closest Bi planes in adjacent quintuple layers was expanded. In doped bulk samples an opposite behavior for Bi-Bi inter and intra-QL spacing was observed. All those results are direct and strong support for suggested structural explanation of qualitative distinction between non-superconducting thin films and superconducting bulk samples found in our work. We suppose, that in bulk crystals Sr atoms are predominantly intercalated in vdW gap, while in thin films dopant atoms reside
also on Bi sites and occupy different interstitial coordination
positions inside quintuple layers.

It is not necessarily that Sr atoms are randomly distributed in the lattice. We can't exclude formation of metal-rich nanoclusters~\cite{JMMM331,Nan27}, flat inclusions~\cite{JPCM30}, segregation~\cite{ACSnano}, or even well arranged blocks~\cite{arx1810}. However, in XRD similar imperfections are usually manifested just as a broadening of the reflections, while detection of the specific Sr-enriched structures by means of TEM is rather challenging. To the best of our knowledge none of the published TEM pictures of SC Sr$_x$Bi$_2$Se$_3$ demonstrate explicitly locations of Sr atoms. Small atomic number of Sr and low concentration (~ 1.2 at. \%) even in SC bulk crystals, along with probable stochastic distribution make TEM imaging of Sr a rather complicated task. Yet eventually, actual dopant housing is a key for understanding their impact on carrier concentration and superconductivity.

\section{\label{concl}Conclusion}
In our paper we have grown a series of Sr-doped Bi$_2$Se$_3$ films on ($1\,1\,1$) BaF$_2$ substrate with various Sr content and thickness. These films are not superconductive and differ strongly from bulk crystals by structure, morphology and transport properties. We explain these differences by various positions, that Sr atoms take in the crystalline lattices: mostly intercalation in crystals and predominantly substitution/interstitial in thin films. These differences are due to different growth temperature ($\sim$1100~K for crystals and $\sim$550~K for films), distinct liquid-solid and vapor-solid crystal formation mechanism, unequally maintained stoichiometry, and domain structure of the films. We believe that the path to superconductivity in epitaxial Sr-doped Bi$_2$Se$_3$ films comes through the adjustment of the Sr-atoms subsystem by co-doping and/or interface engineering to attain a proper Sr atom distribution. Another possibility to achieve superconducting doping structure is to fine tune film stoichiometry through lowering Se to Bi$_2$Se$_3$ flux ratio while using a compound bismuth-selenide effusion source. Therefore, a lacking so far understanding (both theoretical and experimental) where the Sr atoms should and use to sit, is highly demanded.

\begin{acknowledgments}
The work was supported by Russian Science Foundation (Grant N 17-12-01544). Magnetotransport, XRD, and AFM measurements were performed using the equipment of the LPI shared facility center. The authors are thankful to V.M. Pudalov for reading the manuscript.

\end{acknowledgments}

%\nocite{*}
%\bibliography{SashaJAPformatREview}% Produces the bibliography via BibTeX.

\end{document}